\documentstyle[sprocl]{article}

\def\Journal#1#2#3#4{{#1} {\bf #2}, #3 (#4)}

\def\NPB{{\em Nucl. Phys.} B}

\def\CMP{\em Comm. Math. Phys.}

\def\st{\scriptstyle}

\def\be{\begin{equation}}
\def\ee{\end{equation}}
\def\bea{\begin{eqnarray}}
\def\eea{\end{eqnarray}}
\def\equ#1{~Eq.(\ref{#1})}
\def\half{{\st\frac{1}{2}}}
\def\cb{{\bar c}}
\def\gb{{\bar\gamma}}
\def\tr{{\rm Tr}}
\def\pa{\partial}
\def\vev#1{<#1>}

\begin{document}

\title{QUANTIZATION IN THE PRESENCE OF GRIBOV AMBIGUITIES}

\author{M. SCHADEN}

\address{New York University, 4 Washington Place, New York,\\ NY
10003, USA\\E-mail: ms68@scires.nyu.edu} 


\maketitle\abstracts{ The non-perturbative validity of 
covariant BRST-quantization of gauge theories on compact Euclidean
space-time manifolds is reviewed. BRST-quantization is related to the 
construction of a Topological Quantum Field Theory (TQFT) of Witten type on
the gauge group. The criterion for the non-perturbative validity of the 
quantization is that the partition function of the corresponding  TQFT
does not vanish and that its (equivariant) BRST-algebra is free of
anomalies. I sketch the construction of a TQFT whose partition function is
proportional to  the generalized Euler-characteristic of the coset
space $SU(n)_{gauge}/SU(n)_{global}$ with an associated equivariant
BRST-algebra that manifestly preserves translational symmetry. 
Some non-perturbative consequences of this approach are discussed.}

\section{Introduction}
In the best of worlds, restricting a  field theory with first
class constraints to a (gauge fixing) surface uniquely determines the
fields and their conjugate momenta and enables one to proceed with the
quantization of the theory as described  by Dirac\cite{Di50}. As
originally observed by Gribov\cite{Gr78} and proven inevitable under
certain circumstances by Singer\cite{Si78}, this is unfortunately not
the situation one encounters in 
non-abelian gauge theories. Due to the topological structure of the
non-abelian gauge group, the solution to local gauge conditions for
the connection never is unique on compact space-times.  An analogous
situation already occurs if one considers bounded functions on a circle,
\be\label{fs1}
f: S_1 \rightarrow R \ .
\ee
Uniqueness requires that these are periodic and there are therefore
always an even number of  solutions to $f=0$ or none. The fact that
solutions to $f=0$ appear and disappear in pairs when $f$ is
continuously deformed, suggests that one can construct a topological
invariant of the circle that does not depend on the function $f$. Indeed,
\be\label{s1top}
\chi=\sum_{{x_0:f(x_0)=0, f'(x_0)\neq 0}} f'(x_0)/|f'(x_0)|
\ee
where $f'$ denotes the derivative  of $f$, is constant under
continuous deformations of $f$. Accordingly, we can evaluate\equ{s1top} at
a function  $f=V'$ which is the derivative  of a bounded Morse potential
\be
V: S_1\rightarrow R \ .
\ee
\equ{s1top} in this case is just Morse's definition of the 
Euler number $\chi(S_1)=0$ of the circle. The example above can 
be generalized to higher dimensional manifolds\cite{ns83}. It
illustrates several points that will be relevant for the 
non-perturbative quantization of a non-abelian gauge theory:
\begin{itemize}
\item whether or not the solution to $f=0$ is ambiguous depends on the
topology of the manifold (the circle in our simple
example) rather than on the function $f$.
\item one may nevertheless construct a topological characteristic of
the manifold that does not depend on continuous deformations of the
function $f$. It is important to note that the topological
characteristic generally requires that one retains  {\it all}
solutions to $f=0$.  
\item certain topological characteristics of the manifold (such as the
Euler number of a circle) may vanish.
\end{itemize}
The latter case also occurs in the conventional quantization of
non-abelian gauge theories. Fortunately it can be circumvented by
either choosing a different topological characteristic of the
gauge-group manifold {\it or by choosing a different manifold}. 

The first option is difficult to implement in the case of gauge theories. 
The BRST-quantization generically computes the
Morse definition of the (generalized) Euler
characteristic of a manifold. To circumvent the problem, we will choose the 
second option. To see how this works consider again the example above.
We could have restricted ourselves to continuous functions $f$ that
vanish at a certain point (call this point the north pole) of the
circle and have a non-vanishing positive derivative there. Omitting
the contribution from the north pole in the definition\equ{s1top} then
results in a topological characteristic $\chi=-1$   that
no longer vanishes. In effect we have replaced the circle by the closed
interval with boundary conditions and are computing a  (generalized)
Euler characteristic of the closed interval.
The equivariant BRST-construction below
removes certain zero's in much the same manner. By considering these
{\it topological} aspects of gauge fixing we found\cite{Ba98} a
global topological 
obstruction (anomaly) to the quantization of an $SU(n)$ gauge theory on compact
manifolds with non-trivial $\pi_3$ homotopy group\cite{Ba98} and that the
topological characteristic is a constant only in {\it connected}
sectors of the orbit space\cite{Ba96}. Fortunately the orbit space of a torus
with periodic boundary conditions for the gauge fields is {\it
connected} and the homotopy group $\pi_3$ of this manifold is
trivial. 

\section{Gauge fixing, BRST and TQFT}
For these reasons,  simplicity, and also because this is a  physically
interesting case, we consider an $SU(n)$ gauge theory on a compact Euclidean
hypertorus as space-time. Of course we are eventually interested in
the thermodynamic limit where the volume of this torus is taken to be
arbitrary large. Why not consider $R_4$ from the outset? Well,
topological considerations generally apply to compact manifolds and it
would be rather surprising if non-abelian
gauge theories could not be defined  on them. Compact
Euclidean space-time furthermore 
{\it regulates} the infrared behavior of massless theories
in a manifestly gauge invariant fashion. The hypertorus also
offers several practical advantages compared to other compact manifolds. Most
prominently it preserves translations and can be considered in
$d$-dimensions: there are $d$ commuting generators, Fourier analysis
is relatively simple, and one 
can still speak of masses, spectra, etc... Best of all, one expects
the thermodynamic limit to be rather smooth. The torus is also one of
the manifolds that admits physical fermion fields, i.e., quarks, with
{\it anti-periodic} boundary conditions. Surprisingly one {\it has to
} choose {\it periodic} boundary conditions for the connection, if
it couples to quark fields in the fundamental representation\footnote{Twisted boundary conditions for
the gauge fields are admissible only in the {\it absence} of fundamental
fermions\cite{tH79}. An immediate consequence of the periodic
boundary conditions for the gauge fields on a {\it finite} torus is
that the Pontryagin number vanishes: the space of orbits is {\it
connected} (the strong CP-problem does not arise in this case, but the
$U_A(1)$-problem is not easily solved on a {\it finite} 
torus in covariant gauges either).}. Consequently also the
ghost- and antighost- fields are periodic in covariant gauges in this
case. I will only discuss covariant gauges in the following, but the
topological arguments below apply equally well to any gauge condition
that can be continuously deformed to Landau gauge  
\be\label{Landau}
\partial\cdot A =0
\ee
Consider the orbit\footnote{for the sake of brevity, the notation 
used here is not entirely consistent. Generally the fields are
anti-hermitian forms in the adjoint representation of the
Lie-algebra of the group. Euclidean indices and the dependence on the
metric of the compact space-time will usually be suppressed.} 
\be\label{AU}
O_A:=\{A^U = U A
U^\dagger +U d U^\dagger: U(x)\in SU(n)\}\ ,
\ee
 labeled by a connection $A$
on it. A connection on the orbit\equ{AU} that satisfies\equ{Landau}
is a local extremum of the bounded Morse functional
\be\label{morse}
V_A[U]:=\int_{\cal T} \tr A^U\cdot A^U
\ee
on the orbit $O_A$\cite{Zw89}. Generically there are many extrema  and thus many
solutions to \equ{Landau} that form a topological space
\be\label{space}
{\cal E}_{A}:=\{U: \partial\cdot A^U=0\}
\ee
The space ${\cal E}_A$, by construction,  depends only on the {\it orbit}
$O_A$. Topological characteristics of 
${\cal E}_A$ furthermore do not depend on continuous
deformations of the orbit (or, for that matter, the gauge
condition). We can therefore quantize the gauge theory 
within a connected sector of its orbit space by constructing a TQFT
that computes a non-vanishing topological number of the space ${\cal
E}_A$. The
procedure to construct a TQFT whose partition function is the
generalized Euler characteristic of a topological space is well
known\cite{Bi91}. This TQFT for the space ${\cal E}_A$ indeed corresponds
to the conventional (covariant) BRST-quantization of a gauge 
theory\cite{Ba98,Ba96}. It is however readily seen that the
generalized Euler number of the space ${\cal E}_{A}$ vanishes for a
gauge group such as $SU(n)$, because the gauge
condition\equ{Landau} is degenerate with respect to {\it
global} gauge transformations. Thus the manifold corresponding to
global (righthanded) gauge-transformations factorizes and ${\cal E}$
for $SU(n)$ is a manifold with the topological structure 
\be\label{decomp}
{\cal E}_{A}\sim {\cal B}_A \times SU(n)_{global}
\ee
If we assume for the moment (this can be verified) that the generalized
Euler character of a product space  
factorizes into the product of the (generalized) Euler characters of
the individual spaces, 
\be\label{factorize}
\chi(M_1\times M_2) =\chi(M_1)\cdot\chi(M_2)
\ee
then $\chi({\cal E}_A)=\chi({\cal B}_A)\cdot \chi(SU(n))=\chi({\cal
B}_A)\cdot 0$ is at best an indefinite expression, because the Euler
character of the compact group manifold $SU(n)$ vanishes. Since the common
$SU(n)$ factor does not depend on the orbit, we may as well omit it
and construct a TQFT that computes the generalized Euler character of
the coset space
\be
{\cal B}_A\simeq {\cal E}_A/SU(n)\ .
\ee
The construction of the corresponding TQFT below leads to  an {\it
equivariant} BRST-quantization. For an $SU(2)$ gauge theory it was
shown that $\chi({\cal B})$ is odd and therefore does not
vanish\cite{Ba98,Sc98}. In certain cases (as for  an $SU(2)$ theory  on
$S_4$ in the trivial sector of the orbit space\cite{Ba96} ) one can show
that $\chi({\cal B})$ is in fact finite. 
For the moment it suffices to observe that there is {\it nothing
wrong} with computing the generalized 
Euler character of the space ${\cal B}_A$ instead of ${\cal E}_A$ and
that this topological number is constant on a  connected space of orbits. 

\subsection{The equivariant BRST-algebra}
The construction of the TQFT on the gauge group whose partition
function is proportional to the generalized Euler character of a coset
space proceeds along the general principles outlined
in\cite{Bi91}. A similar procedure can also be employed to
covariantly gauge fix a lattice gauge theory with nonabelian structure
group\cite{Sc99a}. Below are the main arguments for an $SU(n)$ 
gauge invariance  of the continuum theory\cite{Ba98,Sc98}.

Under infinitesimal righthanded gauge transformations,
\be\label{varU}
\delta U(x)=U(x)\theta(x)\ ,
\ee
where $\theta(x)\in su(n)$ is an element of the Lie-algebra. We may
always decompose  
\be\label{sep}
\theta(x)=\tilde\theta(x) +\theta\ ,
\ee
into a generator of global
gauge transformations and a generator of (pointed) gauge
transformations in the coset space $SU(n)_{gauge}/SU(n)_{global}$. 
To specify $\tilde\theta(x)$ uniquely one can require that it satisfy 
some (global) condition. Note that this is a {\it linear}
problem that does not suffer from any ambiguities. However, the
condition generally either will not preserve translational invariance {\it or}
will be {\it nonlocal}. One may for instance demand that
$\tilde\theta(0)=0$ at a preferred space-time point. To manifestly
preserve the translational invariance of the model on a compact torus, it is 
preferable to demand 
\be
\int_{\cal T} dx \tilde\theta(x) =0\ .
\ee
The first choice would lead to pointed gauges\cite{mi81}, whereas the
second requires the equivariant BRST-algebra below.

The BRST-algebra of a TQFT of Witten type is constructed by
``ghostifying''\cite{Bi91} the variation\equ{varU}, i.e. one
introduces a nilpotent (anticommuting) BRST-variation $s$, such that
\be\label{brsU}
s U(x)=U(x) (c(x) + \omega)
\ee   
where $c(x)$ and $\omega$ are Lie-algebra valued Grassmann ghosts. To
uniquely decompose\equ{brsU} into global and local ghosts,
we eventually will enforce the {\it non-local} constraint 
\be\label{constraint1}
\int_{\cal T} c(x)=0\ . 
\ee
Nilpotency of the operation $s$ implies that
\be\label{nilU}
s^2 U(x)=0\ \Rightarrow \ s c(x)+ s \omega= -\half [c(x),c(x)] -[\omega ,
c(x)] - \half [\omega,\omega]\ ,
\ee
where $[\cdot,\cdot]$ denotes the commutator graded by the ghost number,
i.e., in \equ{nilU} it is the {\it anti-}commutator since the 
ghost number of both fields is odd. We recognize that the
last term on the RHS of\equ{nilU} is global and that the second term
is just an infinitesimal  rigid gauge transformation of $c(x)$
generated by the ghost $\omega$. Consistency of the
constraint\equ{constraint1} under BRST-variation demands that we also
have
\be\label{constraint2}
\int_{\cal T} s c(x)=0
\ee
when \equ{constraint1} is satisfied. We {\it cannot} therefore choose
$s c(x)=-\half [c(x),c(x)] -[\omega ,c(x)]$, since $[c(x),c(x)]$ in general
does have a constant component even if $c(x)$
doesn't. (It is at this point that we deviate from pointed gauges,
where one demands $c(x=0)=0$, which is consistent with $s
c(x=0)=0$ even for non-abelian groups but breaks manifest
translational invariance.)  Consistency of the
constraint\equ{constraint1} under BRST variation  forces one  to
introduce a global ghost field $\phi$ of 
ghost number two and satisfy\equ{nilU} by the algebra
\bea\label{brsc}
s c(x)&=& -\half [c(x), c(x)]-[\omega , c(x)] -\phi\cr
s \omega &=& -\half [\omega,\omega]+\phi\cr
s \phi &=& -[\omega,\phi]\hfil\ ,
\eea
where the BRST-variation of $\phi$  ensures the nilpotency of
$s$. Evidently the constraint\equ{constraint2} is an equation for 
$\phi$. It is worth pausing at this point of the construction to
note that the somewhat complicated BRST-structure we have begun to
develop is a consequence
of the {\it non-abelian} nature of the gauge group. In the abelian
case all the transformations would have been linear and we obviously
could have removed the global invariance by simply imposing the
constraint\equ{constraint1} without further ado. Since the ghosts
furthermore decouple in the abelian case, the
equivariant construction results in the conventional
covariant gauge-fixing 
for {\it abelian} gauge theories. We will see that it gives something
{\it new} in the non-abelian case and that it is worth pursuing this
avenue of attack.

In addition to the usual doublet of covariant gauge constraints
\be\label{constraint3}
\partial\cdot A^U(x)=0\quad{\rm and}\quad s \partial\cdot A^U=\partial\cdot
D^{A^U}c(x) -[\omega,\partial\cdot A^U]=0
\ee
for $U(x)$ and $c(x)$ we now also have to implement the doublet of
global constraints\equ{constraint1} and \equ{constraint2}. The
Lagrange multipliers for the gauge constraints are the local
Nakanishi-Lautrup field $b(x)$ and its
partner, the anti-ghost $\cb(x)$. They form an
equivariant BRST-doublet
\be\label{doublet1}
s\cb(x)= -[\omega, \cb(x)] + b(x)\qquad\qquad sb(x)=-[\omega,
b(x)]+[\phi,\cb(x)]\ . 
\ee
Note that the anti-ghost $\cb(x)$ transforms under global
transformations generated by $\omega$, because the corresponding
constraint\equ{constraint3} does. This is vital and guarantees
that the effective action we will construct does not depend on
$\omega$. 
The BRST-transformation of the
Nakanishi-Lautrup field $b(x)$ ensures the nilpotency
of $s$. Observe that $b(x)$ is {\it not} annihilated by $s$
in\equ{doublet1}. 

To implement the global constraints \equ{constraint1} and
\equ{constraint2} we require a BRST-doublet of global fields
$\bar\sigma, \sigma$. For
by now familiar reasons this doublet satisfies 
\be\label{doublet2}
s\sigma=-[\omega,\sigma]+\bar\sigma\qquad\qquad
s\bar\sigma=-[\omega,\bar\sigma] +[\phi,\sigma]\ .
\ee
In covariant gauges  it is
finally important that the constant modes of $\cb(x)$ and
$b(x)$ do not couple to the constraints\equ{constraint1}. To avoid
uncompensated 
zero-modes of the anti-ghost
and preserve the BRST-invariance of the model on a finite torus, one
has to impose an additional BRST-conjugate pair of global constraints:
\be\label{constraint4}
\int_{\cal T} \cb(x) =\int_{\cal T} s \cb(x) =0\ ,
\ee
and introduce associated global fields $\gamma$ and $\gb$ of ghost
number $1$ and $0$ respectively. This second doublet of global Lagrange
multipliers  satisfies
\be\label{doublet3} 
s\gb=-[\omega,\gb]+\gamma\qquad\qquad
s\gamma=-[\omega,\gamma]+[\phi,\gb]\ .
\ee

\begin{center} 
\begin{tabular}{|l|r|r||r|r|r|r|r|r|r|r|}\hline 
field&$A^U(x)$ & $\psi(x)\&\bar\psi(x)$ & $c(x)$ & $\cb(x)$ & $b(x)$ &
$\phi$ & $\sigma$ & $\bar\sigma$ & $\gb$ & $\gamma$\\ \hline 
dim&$1$&$3/2$&$0$&$2$&$2$&$0$&$4$&$4$&$2$&$2$\\ \hline 
$\phi\Pi$&$0$&$0$&$1$&$-1$&$0$&$2$&$-2$&$-1$&$0$&$1$\\ \hline 
\end{tabular} 
 
\nobreak\vspace{.2cm}{\footnotesize 
{\bf Table 1.} Canonical dimensions and ghost numbers of the fields.}  
\end{center} 

We have completed the construction of the algebra and field content of
the TQFT. The canonical
dimensions and ghost numbers of the fields is summarized in
Table~1. The symmetry generated by $s$ is nilpotent by construction
\be\label{nilpotency}
s^2=0 
\ee
on any functional of the fields in Table~1. The BRST-algebra above is
one for the gauge group elements $U(x)$ and
the connection $A(x)$ is a
background field that does not transform   
\be
s A(x)=0\ .
\ee
Note, however, that
\be\label{rel1}
s A^U(x)=-D^{A^U} c(x)+ [\omega, A^U] \ ,
\ee
where $D^A$ is the usual covariant derivative of the
adjoint representation. \equ{rel1} implies a tight relation between
the BRST-symmetry of the TQFT and the BRST-symmetry of the
corresponding gauge theory. 

\subsection{The TQFT}
The partition function, $Z$, of the TQFT in principle could depend on
the background 
connection $A$ and a set of (gauge) parameters $\{\alpha\}$. Formally
the path integral representation of $Z$ is of the form 
\be\label{Z}
Z[A(x), \{\alpha\}]=\int
{\cal D}[c(x),\cb(x),U(x),\phi,\sigma,\bar\sigma,\gamma,\gb]\ \exp \{ S\}\ , 
\ee
where ${\cal D}[\dots]$ is a BRST-invariant measure and the effective
action $S$ 
\be
 S = s W[c(x),\cb(x),U(x),\sigma,\gb;A(x),\{\alpha\}] 
\ee
is BRST {\it exact}, that is $s$ is a nilpotent BRST-symmetry and $S=s
W$.  Apart from being exact,
the effective action, however, has to satisfy several additional 
criteria in our case.
Note that we {\it do not integrate} over the constant ghost $\omega$
in\equ{Z} and that the action therefore $S$ should not depend on this field
either. Since $\omega$ in \equ{brsc},\equ{doublet1},\equ{doublet2} and
\equ{doublet3} generates {\it global} righthanded gauge
transformations of all the fields {\it except} itself, $S$ is
$\omega$-independent only if $W$ is a globally gauge invariant
functional of the fields that does not depend on $\omega$. It
immediately follows that $W$ can depend on the background one-form $A(x)$ and
$U(x)$ only in the combinations $A^U$ and $U^\dagger dU$ that
transform in the adjoint. In order to use the TQFT as a gauge-fixing
device we  consider actions $S$ that depend on $U$ only via
$A^U$.

In addition the effective action $S$ should
preserve ghost number, be manifestly $SO(4)$ invariant (or at least
invariant under the discrete hypercubic subgroup 
of the hypertorus) and power counting
renormalizable in four dimensions. This requires that we construct a
manifestly $SO(4)$- and globally gauge invariant local functional $W$ of
ghost number $-1$  and canonical dimension $4$ from the fields of
Table~1.
Up to arbitrary normalizations of the fields and constants, the most
general power-counting renormalizable action $S$ with the 
required symmetries thus is the BRST-variation of
\bea
W=&-2 \int_{\cal T}\tr\left[\cb(x)\pa\cdot A^U(x) +\half\alpha \cb(x) b(x)
+\alpha\delta\cb(x)\cb(x) c(x)\right.\cr
&\qquad\left. +\alpha\rho\cb(x) A^U(x) A^U(x) -\gb\cb(x)
-\sigma c(x)\right]\ ,
\eea
which for $SU(n>2)$ depends on {\it three} gauge parameters:
$\alpha,\delta$ and $\rho$. For $SU(2)$, $\rho$ is an irrelevant
parameter because it multiplies terms proportional to the symmetric
structure constants $d^{abc}$. 

No matter how complicated the general covariant effective action
 finally is, the most important point at this stage is that $S$ is BRST-{\it
 exact}. One can prove\cite{Ba98} that the equivariant BRST-algebra we derived is
 free of anomalies for a hypertorus as spacetime. Once this  is established, it
 is not difficult to show that $Z[A(x),\alpha,\delta,\rho]$ in fact
 {\it does not} depend on variations of its parameters,
\be\label{constancy}
\frac{\delta Z}{\delta A(x)} =\frac{\pa Z}{\pa\alpha}=\frac{\pa
 Z}{\pa\delta}= \frac{\pa Z}{\pa\rho}=0 \ ,
\ee
since these variations are proportional to expectation values of
 $s$-exact functionals. This  establishes that\equ{Z} is a
 gauge-parameter independent  constant on a connected sector of the
 parameter and orbit space.

The interesting question is whether this constant is {\it
 normalizable} or whether $Z$ vanishes identically. One only has to
 prove that $Z$ is normalizable (or not) for at least {\it
 one} orbit $A(x)$ and some (admissible) value of the gauge
 parameters. For simplicity we will only consider gauges with
 $\delta=1,\rho=0$ in the following. For this choice of the
 gauge parameters one obtains the simplest effective action 
\bea\label{simpleS}
\left.S\right|_{\delta=1,\rho=0} &= - 2 \int_{\cal
 T}\tr\left[b(x)\pa\cdot A^U(x) -\cb(x)\pa\cdot 
 D^{A^U} c(x)\right.\cr
&\qquad\left. + \alpha\{\half b^2(x)+ b(x)[\cb(x),c(x)] -\cb^2(x)
 c^2(x)\}\right.\cr
&\qquad \left. -\gb b(x) -\gamma\cb(x) -\bar\sigma c(x) +\sigma c^2(x)
 +\sigma\phi\right]\ . 
\eea
This  action of the TQFT  can be further simplified by using the equations
 of motion of some of the fields. The equation of motion of the
 topological ghost $\phi$ sets $\sigma=0$ and those of
 $\bar\sigma$ and $\gamma$ eliminate the {\it constant} modes of the
 ghost and anti-ghost fields on the torus. Finally, the equation of
 motion of the 
 Nakanishi-Lautrup field $b(x)$ is 
\be\label{emb}
b(x)=\gb -\alpha^{-1}\pa\cdot A^U(x)-[\cb(x),c(x)]\ .
\ee
We can thus rewrite $\tilde
 Z[A]:=Z[A(x),\alpha,\delta=1,\rho=0]$ as the relatively transparent
 functional integral  
\be\label{tZ}
\tilde Z[A]={\cal N}(\alpha)\int d\gb \int {\cal
 D}^\prime[c(x),\cb(x),U(x)] \exp 
 S_{GF}[c(x),\cb(x),A^U(x),\gb;\alpha]\ , 
\ee
with an $\alpha$-dependent normalization ${\cal N}$ and a measure
 $D^\prime$ of the dynamical fields  that {\it does not} include
 constant modes of the ghost 
 and anti-ghost. The effective action in\equ{tZ} is
\bea\label{Seff}
S_{GF}[c(x),\cb(x),A(x),\gb;\alpha] &= 2 \int_{\cal T}\tr\left[\left(\pa\cdot
 A(x)\right)^2/(2\alpha) +\cb(x) D^A\cdot\pa c(x)\right.\cr
&\qquad\left. -\gb [\cb(x),c(x)] +
 \gb^2/(2\alpha)\right]
\eea
after dropping a surface-term $\int_{\cal T} \tr\gb \pa\cdot A(x)$ that
vanishes on a torus with periodic boundary conditions for the
 connection.

Note that the effective action\equ{Seff} is unbounded below for
 $\alpha<0$. This does {\it not}
 contradict\equ{constancy} and $\tilde Z$ does not depend on 
 the value of $\alpha$ in the {\it open} interval $\alpha\in
 (0,\infty)$. Landau-gauge is, however, can only be defined as the limit
 $\alpha\rightarrow 0_+$ 
 (which generally is not equivalent to simply {\it setting} $\alpha=0$
 in\equ{simpleS}). 

\section{Why deal with an equivariant BRST-algebra?}
Apart from an interchange of the ghost and
 anti-ghost, the action\equ{Seff} we have 
 constructed is that of conventional covariant gauges when we set
 $\gb=0$. There is one other important difference: the measure ${\cal
 D}^\prime$
 of the functional integral\equ{tZ} no longer includes an integration over
 constant ghost and anti-ghost modes. The importance of these
 differences lies in the fact that one can show that the TQFT for conventional 
covariant gauge fixing, described by  a partition function 
\be\label{Zconv}
Z_{conv.}[A,\alpha]:=\int D[c(x),\cb(x),U(x)]\ \exp
 S_{GF}[c(x),\cb(x),A^U(x),\gb=0;\alpha]\ , 
\ee
vanishes for $\alpha>0$ on any finite hypertorus whereas\equ{tZ} is
 normalizable\cite{Ba98,Sc98}. The argument that\equ{Zconv} vanishes
relies on the fact that the effective action of\equ{Zconv} is 
 BRST-exact with respect to the on-shell nilpotent BRST-symmetry
 $\tilde s$ defined by
\be\label{convbrs}
\tilde s U(x)=U(x)c(x),\ \tilde s c(x)=-c^2(x)\ 
\tilde s \cb(x)=-\alpha^{-1}\pa\cdot A^U(x)-[\cb(x),c(x)]\ .
\ee
This algebra is nothing but the equivariant BRST-algebra generated by
 $s$ on-shell (i.e.,  when\equ{emb} holds) and all the global fields
 are formally set to zero. The slightly unusual transformation of the
 anti-ghost in\equ{convbrs} just interchanges the role
 of ghost and anti-ghost in the effective action. The argument below
 can also be repeated with the more conventional action and
 corresponding BRST-symmetry. The main point is that a nilpotent BRST-symmetry
 also exists for the TQFT corresponding to conventional covariant
 gauges. It guarantees that $Z_{conv.}$ also {\it does  not} depend on
 on the connection. It thus suffices to show that $Z_{conv.}$
 vanishes for a particular orbit. We choose the trivial orbit
 represented by the connection $A(x)=0$. This is an example of a
 degenerate orbit, since 
\be\label{degenerate}
D^A \eta(x)=0
\ee
has a non-trivial solution $\eta(x)\neq 0$. Other examples of
degenerate orbits  are those corresponding to non-abelian monopole
 configurations, for which the connection is invariant under a 
 $U(1)$ subgroup of $SU(n)$ that defines the magnetic charge of the
 monopole. 

To appreciate the significance of degenerate orbits note that
 \equ{degenerate} implies 
 that the connection $A$ is invariant under a particular
 (infinitesimal) gauge
 transformation generated by $\eta(x)$ (hence is ``degenerate''). 
 Since the definition\equ{AU} together
 with\equ{degenerate} shows that
\be\label{zero}
D^{A^U} (U(x)\eta(x) U^\dagger(x))=U(x)[ D^A\eta(x)] U^\dagger(x)=0\ ,
\ee
for any gauge transformation $U(x)$, it furthermore does not matter which
 connection we pick to represent the degenerate orbit. We can
 evidently find a gauge 
 transformation $U_0(x)$ and a corresponding connection
 $A_0(x)=A^{U_0}$ on the degenerate orbit for which
 $\eta_0(x)=U_0(x)\eta(x)U_0^\dagger(x)$ is diagonal. For $\eta_0$ in
 the Cartan subalgebra\equ{zero} requires that
\be\label{diag}
d \eta_0(x)=0\quad{\rm and}\quad [A_0(x),\eta_0(x)]=0\ ,
\ee   
separately\footnote{I am indebted to D. Zwanziger for this argument},
 since the commutator is not diagonal. The zero-form $eta_0$
thus does not depend on  space-time and the connection $A_0(x)$ is in
a sub-algebra of $su(n)$. The mode $\eta_0$ is  normalizable on
 {\it any} compact Euclidean manifold such as a
 hypertorus. Since $\eta(x)$ and $\eta_0$ are related by a unitary
 transformation, we furthermore have that
\be\label{norm}
 \tr\eta^\dagger(x)\eta(x)=\tr\eta_0^\dagger\eta_0= const.
\ee
is independent of the space-time point $x$. \equ{norm} not only implies
 that the modes satisfying\equ{degenerate} are normalizable on a
 compact space-time manifold  for any
 connection of the degenerate orbit, but also that there is {\it no
 solution} to\equ{degenerate} in the space of functions $\eta(x)$
 satisfying the condition $\eta(0)=0$. Recalling that $\eta(x)$
 is the generator of an infinitesimal gauge transformation, the
 restriction to $\eta(0)=0$ is evidently equivalent to considering
 only gauge transformations of the {\it pointed} gauge group
 $SU(n)_{gauge}/SU(n)_{global}$ rather than those of
 $SU(n)_{gauge}$. Thus there are {\it no degenerate 
 orbits} with respect to the {\it pointed} gauge group for which we
 constructed the {\it equivariant} BRST-algebra. On the other hand,
 there are degenerate connections with respect to the full gauge
 group. Since $\eta(x)$ satisfying \equ{degenerate} is also a
 zero-mode of the Faddeev-Popov (FP) operator $\pa\cdot D^A$, degenerate
 orbits are {\it on} the Gribov horizon, that is $\det \pa D^A=0$ for
 {\it any} connection on a degenerate orbit and in particular for all
 those connections of the degenerate orbit that satisfy a covariant
 gauge condition. 
Thus there are normalizable ghost and anti-ghost zero-modes on any
 degenerate orbit, leading to a vanishing FP-determinant on the {\it
 whole} orbit. One can also evaluate $Z_{conv.}[A(x)=0,\alpha]$
 semiclassically by choosing the gauge parameter $\alpha$ sufficiently
 small and show more 
 rigorously that $Z_{conv.}$ is proportional to $\chi(SU(n))$ and
 indeed vanishes\cite{Ba98,Sc98}.

On the other hand $\tilde Z$ defined by\equ{tZ} does not suffer from
Grassmannian zero-modes. The modified FP-determinant  $\det [D^A\cdot\pa
+\hat\gb]$ generically does not vanish on a degenerate orbit. Constant
ghost modes have been eliminated and the determinant depends on the
global ghost $\gb$. By evaluating $\tilde Z$ semiclassically, one
can prove that the partition function of the TQFT is normalizable.

Because $\tilde Z$ does not depend on the orbit and does not vanish,
one can insert $\tilde 
Z$  in the functional integral of the gauge theory and perform the 
change of variables
\be\label{changevar}
A^U\rightarrow A
\ee
to decouple the integration over gauge transformations in each
connected sector of the orbit space\cite{Ba98}. The change of
variables\equ{changevar} implies that the gauge-fixed theory
inherits the equivariant BRST-symmetry of the TQFT. Physical observables of
this theory are expectation values of {\it globally} gauge-
and BRST-invariant functionals with vanishing ghost number. 

\subsection{The role of the global ghost $\gb$}
In addition to the usual dynamical fields, the equivariantly
quantized non-abelian gauge theory also depends on the global ghost
$\gb$. As far as the dynamical fields are concerned, this global field is a
somewhat unconventional ``mass''-parameter for the ghosts. One can
compute the effective measure for this moduli-space order-by-order
in the loop expansion for the dynamical fields. For $n_f=n$ quark
flavors it was shown\cite{Sc98,Sc98b} that there is a nontrivial ultraviolet fixed point  
\be\label{expect}
\vev{\gb}|_{g\rightarrow 0, \mu\rightarrow\infty, 
 \Lambda_{MS}(\mu,g) {\rm fixed}}\neq 0
\ee
in the thermodynamic limit of the hypertorus. For $n_f<n$ the
existence of such a fixed point 
in the more complicated moduli-space with $\delta\neq 1$ was also
established\cite{Sc98}. Although\equ{expect} implies that the global gauge
symmetry of the equivariantly gauge-fixed theory is broken, the
corresponding Goldstone poles do {\it not} appear in  {\it
observables} of the equivariant cohomology of the model, since these
are globally gauge invariant (just as th Goldstone poles of
spontaneously broken  chiral symmetry  are not found in
chirally invariant correlation functions). This is not to say that the
corresponding Goldstone singularities in {\it unphysical} correlators
have no effect--they could even be the cause for confinement in
covariantly quantized gauge theories\cite{Sc98}. \equ{expect} furthermore
{\it does not} imply that the equivariant BRST-symmetry is broken, if
the {\it gap} equation 
\be\label{gap}
\vev{s\cb(x)}=0=\vev{\gb} -\vev{[\cb(x),c(x)]}\ ,
\ee
is satisfied. \equ{gap} is a consequence of\equ{doublet1},\equ{emb} and
translational invariance.   Since a renormalization group invariant
minimum of the effective 
potential on the moduli-space is just another definition of the
asymptotic scale parameter of the theory, the expectation
value\equ{expect} is related to $\Lambda_{MS}$ by a one-loop
calculation. $\vev{\tr\gb^2}$ furthermore is the expectation value of the
scale anomaly when all dynamical fields have been integrated out. The resulting
relation between the expectation value of the energy-momentum tensor
of $SU(n)$ gauge theory with $n_f=n=3$ quark flavors  and the
asymptotic scale parameter  compares 
favorably with phenomenological estimates from QCD sumrules\cite{Sc98b}.
Power corrections to the asymptotic behavior of physical
correlation functions proportional to
$\tr{\gb^2},\tr{\gb^3},\dots$etc. arise order by order in the loop
expansion for the dynamical fields when\equ{expect} holds, even in the
chiral limit of a massless theory. Note the absence of dimension $2$
contributions, a consequence of the fact that physical correlators are
globally gauge invariant and $\gb$ is a field of canonical dimension $2$ in the
adjoint. The power corrections therefore have the structure required  by 
Wilson's Operator Product Expansion for a gauge theory without
scalars. It will be interesting to compare the two asymptotic
expansions on a more quantitative level.  

\section*{Acknowledgments}
I would like to use this opportunity to thank the organizers for a very
interesting, lively and informative workshop. This work would not have
been possible without the continuing and active support by
D.~Zwanziger and L.~Spruch. I am also indebted to L.~Baulieu and the 
LPTHE for support.

\end{document}